\documentclass[
]{ceurart}

\usepackage{listings}
\usepackage{csquotes}
\usepackage{subfig}
\begin{document}

\copyrightyear{2023}
\copyrightclause{Copyright for this paper by its authors.\newline
  Use permitted under Creative Commons License Attribution 4.0
  International (CC BY-NC-SA 4.0).}

\conference{BIR-WS 2023: BIR 2023 Workshops and Doctoral Consortium, 22nd International Conference on Perspectives in Business Informatics Research (BIR 2023), September 13-15, 2023, Ascoli Piceno, Italy}

\title{Constraint based Modeling according to Reference Design}


\author[1]{Erik Heiland}[%
orcid=0000-0002-6636-8356,
email=erik.heiland@unibw.de,
]
\cormark[1]
\fnmark[1]
\address[1]{Universität der Bundeswehr München,\\ Werner-Heisenberg-Weg 39, 85577 Neubiberg, Germany}

\author[1]{Peter Hillmann}[%
orcid=0000-0003-4346-4510,
email=peter.hillmann@unibw.de,
]
\cormark[1]
\fnmark[1]

\author[1]{Andreas Karcher}[%
email=andreas.karcher@unibw.de,
]
\cormark[1]
\fnmark[1]

\cortext[1]{Corresponding author.}
\fntext[1]{These authors contributed equally.}

\begin{abstract}
Reference models in form of best practices are an essential element to ensured knowledge as design for reuse. Popular modeling approaches do not offer mechanisms to embed reference models in a supporting way, let alone a repository of it. Therefore, it is hardly possible to profit from this expertise. The problem is that the reference models are not described formally enough to be helpful in developing solutions. Consequently, the challenge is about the process, how a user can be supported in designing dedicated solutions assisted by reference models. In this paper, we present a generic approach for the formal description of reference models using semantic technologies and their application. Our modeling assistant allows the construction of solution models using different techniques based on reference building blocks. This environment enables the subsequent verification of the developed designs against the reference models for conformity. Therefore, our reference modeling assistant highlights the interdependency. The application of these techniques contributes to the formalization of requirements and finally to quality assurance in context of maturity model. It is possible to use multiple reference models in context of system of system designs. The approach is evaluated in industrial area and it can be integrated into different modeling landscapes.
\end{abstract}


\maketitle

\section{Introduction}
\label{sec:intro}
Reference models (RM) are used in many areas today and are becoming more and more popular. 
The complexity of systems and their increasing interconnection require a high degree of knowledge from various areas.
Only a structured processing of this knowledge as well as the consequent further development of this enables rapid technological progress.
Thus, most companies business initially rely on structures and processes that have proven themselves in general, and later continuously adapt them to their needs.
These generally accepted facts can be called RM and can be found almost everywhere, also in collections~\cite{Hillmann2022}.
In general, we understand a reference model to be a collection of generally accepted recommended solutions to a specific problem.
IT service management concepts, for example, are characterized by frameworks such as the \textit{IT Infrastructure Library (ITIL)}, IT4IT~\cite{IT4IT}, and within modeling languages like ArchiMate~\cite{Archimate32}, which are used by providers as a basis for orientation.
In the field of vehicle development and production design, the companies also benefit from years of experience.
Industrial reference models such as RAMI~4.0~\cite{RAMI} or the Standard for the Exchange of Product Model Data (STEP) provide standardized interfaces for compatibility and orientation for further development in the context of digitalization.
As last example we want to mention enterprise architectures.
The continuous alignment of business and IT is one of the main challenges of today's companies to remain competitive.
The use of RM or entire frameworks is almost indispensable from a certain size of company on in order to meet this challenge.

The use of RM provides several advantages for organizations.
One example is the service of federated identity management~\cite{Poehn2021}.
First, the reuse of already proven structures and methods saves time and costs. 
Furthermore, the use of RM contributes to quality assurance and standardization.

RM are characterized by universality and reusability within its scope.
The typical process of reuse consists of the adaptation, adoption and extension of the models, and differs depending on the design technique. 
Popular techniques are here for example the construction by configuration or the use of RM by aggregation.

It is not always easy to find the right RM for the specific application context.
Sometimes models are constructed by using several RM, each of which describes only a partial aspect. 
In this case the term \textit{Building Blocks} is usually used.
After the construction of the solution model, another challenge is the comparison with the RM. 
It should not only be used as a modeling aid, but also for quality assurance and compatibility. 
Therefore, it must be traceable to what extent the recommendations of the RM have been implemented.

Such mechanisms are typically not found in commonly used modeling languages and supporting tools for enterprise description, but some approaches based on ontologies respectively knowledge graphs already exist~\cite{Laurenzi.2018, Glaser.2022, Karagiannis2018APF, Mihai.2019, Oliver.2009}.
The challenge is to find a general method to transform arbitrary models into such graphs and to adapt mechanisms of reasoning and querying to the validation of models with respect to reference models.

In this paper we present a generic approach for the formal description of RM that can be integrated into different modeling landscapes.
Our reference modeling assistant allows the construction of application models using different techniques and checks the resulting solutions against the RM for conformity.
Thus, it combines best practices of reference modeling and ensures sustainability.
The application of these techniques contributes to the formalization of requirements and finally to quality assurance in modeling.


\section{Scenario and Requirements}
\label{sec:problem}

The challenges of using RM is illustrated by a simplified example from the automotive industry.
We design a RM for the development of vehicles.
Similar configuration tools can be found on most of the car manufacturers' websites, allowing customers to customize their new vehicle.
Our RM in this scenario contains the following statements, which should apply to every vehicle (regardless of whether the statements are actually correct):
\newpage
\begin{enumerate}
	\item Each car has 4 wheels.
	\item A vehicle has an automatic or manual transmission.
	\item The car has at least one engine. It can have one combustion engine that runs on diesel, petrol or gas. In addition, it can be equipped with an electric motor (hybrid car) or it can be powered exclusively by one or more electric motors. 
	\item If the car is equipped with a combustion engine, it needs a fuel tank, and if it has an electric motor, it needs a battery to store energy.
	\item Vehicles with electric motors can only be configured with automatic transmission.
\end{enumerate}

\begin{figure}[]
	\centering
	\includegraphics[width=0.6\linewidth]{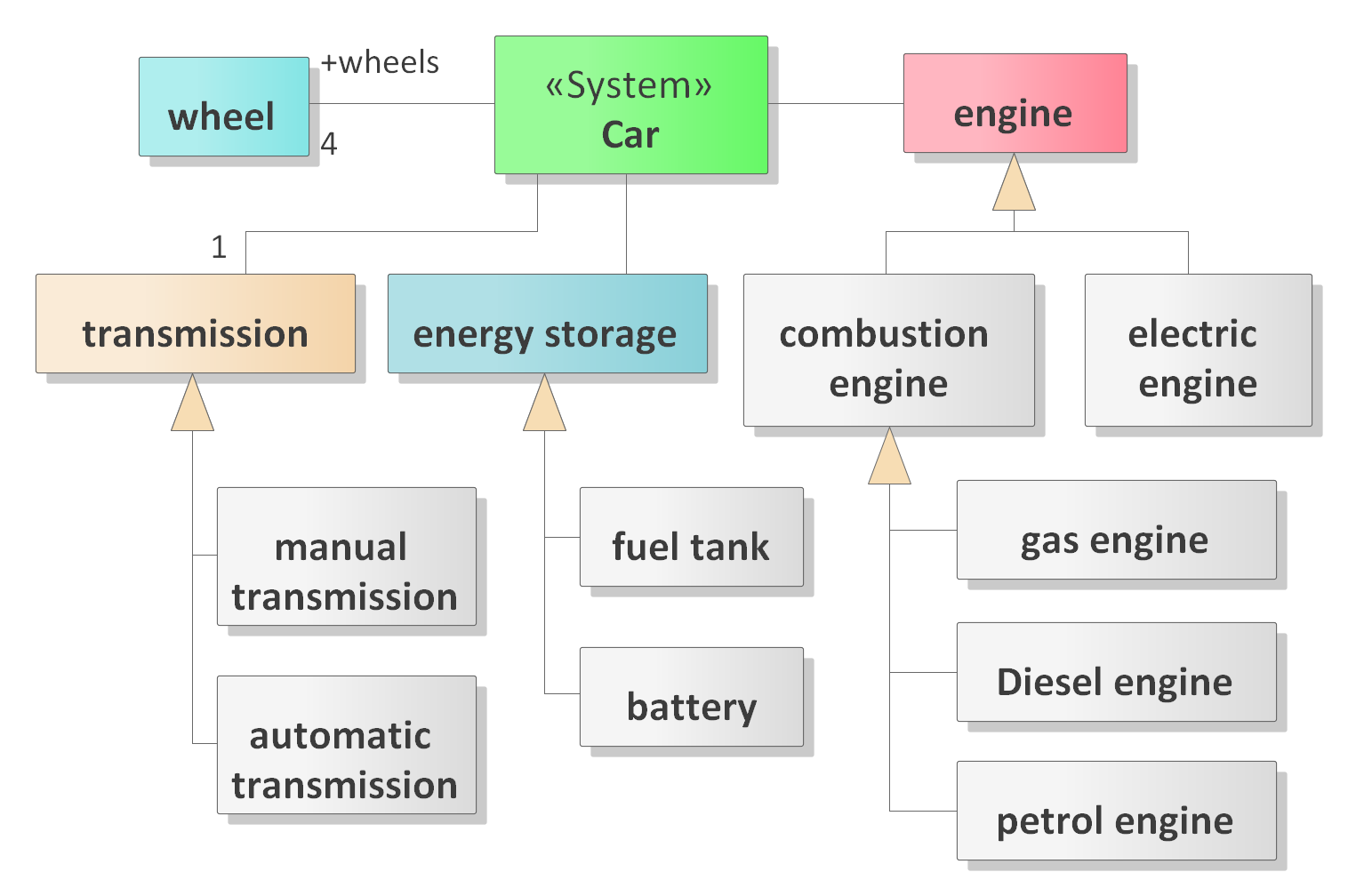}
	\caption[]{Example reference model for car configuration}
	\label{fig:car}
\end{figure}

Figure~\ref{fig:car} shows an example visualization of the RM with its main components as an UML diagram.
This RM can be described as an abstract configuration model in which the user can choose only between these elements.
This model can be further enriched with instances for these classes, where a specific selection of such a component is possible.
The challenge is now the application of this RM by the user, who wants to model a new vehicle based on these specified building blocks.
The question arises how to ensure that all rules of the RM are followed during modeling.
In addition, it must be checked which model aspects are relevant for the user and whether there exist already other solutions from this area that the user could use.

It is noticeable that the RM does not contain all the requirements described above.
Conditions 4 and 5 are not illustrated, but only described in text form.
Even if these could have been formulated in UML, it is often not described formally enough to be supported within the modeling. 
Constraints are mostly supplemented by textual descriptions (e.g. automated processing).
Furthermore, the question arises how the RM can actually support the user in the design of new models.
Usually, the process of reuse consists of taking the results of a RM and then being able to modify and extend them~\cite{Becker.2004}.
Popular modeling tools like ARIS, Enterprise Architect Sparx, Visual Paradigm do not offer such mechanisms to embed RM in a supporting way.
It is therefore hardly possible to profit from the RM advantages mentioned in Section \ref{sec:intro}.

The lack of support in the application of RM has several reasons.
The multitude of heterogeneous modeling tools and languages makes it difficult to implement a universal approach.
In addition, there are different ways of using RM. 
Besides strongly restrictive techniques such as configuration, the instantiation or placeholder technique allows for example some degrees of freedom in the application~\cite{Becker.2007b}.
In comparison, RM only have a recommending aspect and are therefore less restrictive than metamodels, which is why deviations must also be allowed.
In summary, there is a lack of methodology for describing conditions associated with an RM.

Such problems lead to the fact that overarching RM are not formally described, but rather in natural language in the form of guidelines and illustrations.
It is left to the user to interpret the contents correctly and to implement them in his modeling environment.
In the end, it lacks user support for the use of the RM as well as verification mechanisms to ensure the correctness of the solution model.


\section{Related Work}
\label{sec:related_work}

One of the most popular languages for defining constraints for models is the \textit{Object Constraint Language (OCL)}~\cite{OCL24}.
The latest version 2.4 of OCL can be used in conjunction with UML 2.4.1 and MOF 2.4.1.
For models, the language can be used to create invariants, pre- and postconditions for models, among others.
The creation of OCL constraints and syntax checking is supported by most modeling tools. 
Support is provided for example by \textit{USE}\footnote{https://sourceforge.net/projects/useocl/} and \textit{Eclipse OCL}\footnote{https://projects.eclipse.org/projects/modeling.mdt.ocl}.
Apart from the fact that OCL is primarily designed for use with UML, the corresponding support for the application of the RM is missing.
This means that the user cannot be provided with classes and instances as recommendations, which is an essential part of our approach.
Furthermore, someone using UML could define requirements that require a specific stereotype for a certain connection or element. 
This would not help a user who creates their models with ARIS or ArchiMate, in which the concept of stereotype does not exist.

The book~\cite{Fettke2006} provides a motivation for reference modeling and the benefits are shown.
A description form and methodology for the configuration under consideration of constraints does not take place.
However, the paper~\cite{mci/Frank2007} described the notion of open reference models based on analogies to the development of open source software.
Beyond that no detailed description for a solution approach takes place.

Another possibility to realize content-specific restrictions is the development of a \textit{Domain Specific Language (DSL)}.
Tools such as \textit{MetaEdit+}\footnote{https://www.metacase.com} or \textit{ADOxx}\footnote{https://www.adoxx.org/} allow the addition of rules, which can create restrictions for the construction of the models.
However, this approach is based on meta-modeling and is not comparable to the concept of reference modeling.
Our considered problem requires the RM and the solution model in parallel at equivalent levels with a common metamodel.

In the field of process modeling, aggregation and hierarchy are common techniques for reusing RM. 
For example, the modeling languages ARIS and BPMN offer the possibility to model processes that are referenced in other models by using only one aggregated element as inheritance that contains the corresponding process logic. 
Various process management and workflow systems such as~\cite{SignavioGmbH.2020} and~\cite{CamundaInc..} exist for this purpose.
Although this provides support for the use of RM as aggregation, but the application context here is limited to processes.

This work~\cite{Allemang2005} with OWL offers a possibility with the focus on the representation of models.
Semantic Web technologies are also used in the process.
However, it allows only the consistent extension of existing RM.
Dependencies within models for configuration are not considered.

The approaches~\cite{Hooshmand2017,Ascher2022} describes a uniform methodology for describing reference designs.
By including requirements, a continuous modeling and traceability is shown, but dependencies within the model cannot be realized.

Specific instances of RM can be made via a configuration~\cite{Becker2008}.
However, this approach only addresses the reference modelling to an overlying ontology and taxonomy.
Dependencies and conditions within a model are not considered.

Further construction techniques as surveyed in~\cite{Esswein.2016} are not or only insufficiently supported by modeling languages.
In summary, non of the current approaches allows a retrieval of suitable building blocks, their combination and the immediate validation of the solution.




%
%

\section{Concept of Reference Model Design and Application Methodology}
\label{sec:concept}
In this section, we describe our generic approach to constrained-based design of RM.
We show how requirements can be formalized and how solution models can be derived from RM and checked against them.
Our approach is based on a closed world view to obtain a structural and functional connection.

\begin{figure}
	\centering
	\includegraphics[width=.6\linewidth]{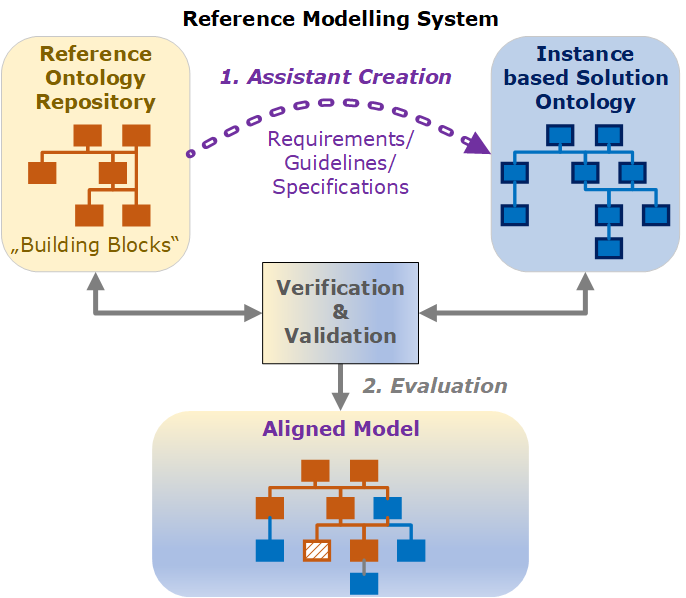}
	\caption{Concept Overview}
	\label{fig:concept}
\end{figure}

An overview of the general principles is shown in Figure~\ref{fig:concept}. 
The concept is to represent the RM and the associated solution models using the \textit{Resource Description Framework (RDF)}~\cite{Cyganiak:14:RCA}.
RDF is chosen for general applicability, especially compatibility with web technologies.
In this form, semantic technologies can be applied to the RM.
This includes the application of rules, reasoning and the use of powerful query languages.
Based on the RM, constraints can be formulated that are based on concepts of predicate logic.
Using the provided reference ontology, solution models can be created.
Finally, SPARQL~\cite{Seaborne:13:SQL} is used as the query language for the validation of the models.
Other languages such as SHACL~\cite{Kontokostas:17:SCL} or SWRL~\cite{Horrocks2004} can also be an alternative to formulate and check the validity of rules based on RDF models. 
The following four aspects of the approach will be described in detail:
\begin{enumerate}
	\item Reference Ontology Design
	\item Constraints Definition
	\item Solution Building
	\item Model Verification and Validation
\end{enumerate}

\subsection{Reference Ontology Design}
In this first stage, we form the basic structure of a RM.
It consists of the creation of a RDF graph with the elements, connectors and attributes 
relevant for the RM.
We further use the vocabulary of \textit{RDF-Schema (RDFS)}~\cite{Haudebourg:23:RS}, especially for describing classes and properties and the generation of taxonomies using \textit{rdfs:subClassOf}.
The result can be regarded as a lightweight ontology.
We consider the case that a RM is to be described in the same modeling language as the solution models.
Therefore, a mapping or a transformation from this language to RDF is necessary.
However, this step is not mandatory for the RM if it is developed directly as an RDF graph.

\begin{figure}[htb]
	\centering
	\includegraphics[width=.6\linewidth]{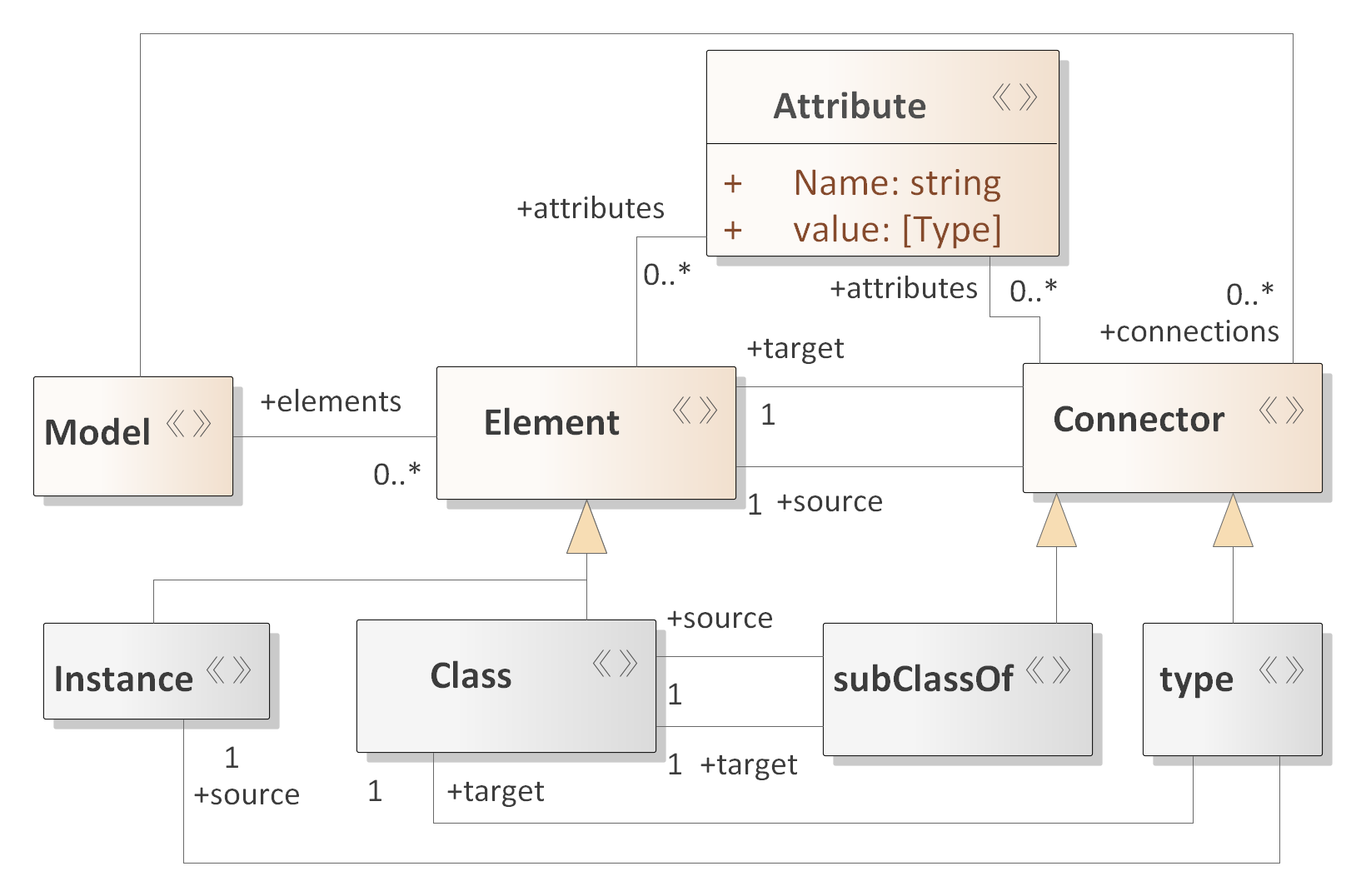}
	\caption{Metamodel for Reference Model Definition}
	\label{fig:referencemodeldefinition}
\end{figure}

For such a mapping, we have used a basic metamodel with essential concepts, which should be valid for most modeling languages.
Thus, a model consists of a set of elements that have properties and are connected to each other via connectors. For the specification of a schema elements of the type "Class" are used, which can be related by means of the relation "subClassOf". Classes are used to represent the reference model. Derived solution models or single solution modules are described with the help of instances, which in turn are typed with classes of the RM.
A representation of the metamodel as a UML profile is shown in Figure~\ref{fig:referencemodeldefinition}.

For each RM a graph needs to be created first. 
A URI node is automatically generated for each element contained in it. 
Each connector is also represented by a URI node. 
Furthermore, it is defined as a property in an RDF triple, which relates the source and target node. 
Element and connector attributes are created as RDF literals and assigned by triples to the corresponding nodes.
Especially for the context of reference modeling, the elements are now distinguished into classes and instances. 
Classes can be hierarchically related via the subclass relationship and instances can be typed with classes. 
This distinction is common in the design of ontologies and helps us to differentiate between abstract placeholders and concrete configuration elements in reference modeling.
A hierarchical arrangement for properties is also possible. 
Optionally source and target elements of the relations can be defined using \textit{rdfs:domain} and \textit{rdfs:range.} 

If the graph is generated from an existing model as in our example, it has to be defined individually how the distinction between classes and individuals is detected. 
This can be done by using certain annotations on the model.
For our example we use the standard UML relation \textit{\enquote{instantiate}} to recognize individuals and \textit{\enquote{generalization}} to build a taxonomy using \textit{rdfs:subClassOf}.

\begin{figure}
	\centering
	\includegraphics[width=.7\linewidth]{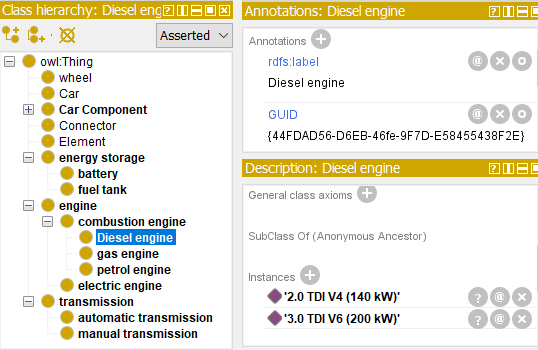}
	\caption[]{Generated taxonomy for the car reference model}
	\label{fig:cargraph}
\end{figure}

Each element and connector has a URI by which they can be identified.
The attributes can be defined as RDF literals.
A section of the resulting taxonomy from our vehicle example is shown in Figure~\ref{fig:cargraph}, implemented with the tool \textit{Proteg\'e}.
This also shows that the RM can already contain concrete instances for the different classes, which can be used for the configuration of the solution. 
In this example two types of diesel engines exist.

\subsection{Constrained Definition}

The second step is to define the constraints.
A distinction is made between structural and value-based constraints.
Structural constraints refer to the existence of certain elements, relationships and attributes.
Complex model relationships across several elements can thereby also be expressed as conditions.
An example for a structural constraint could be: 'Each car requires at least one engine.'
The conditions described in our example are all structural constraints.

Value-based constraints, on the other hand, refer to specific attributes of elements and connectors.
An example could be: 'The sum of the weight of all car components must not exceed 2.500 kg'.
This is also used to ensure appropriate interfaces between elements.
It is in line with the general reference modeling methodology for building blocks and pattern~\cite{Ascher2022}.

Both types of constraints are achieved by comparing values obtained by SPARQL queries from the solution model.
For structural requirements a number of existing elements or paths is usually determined. 
This can be done with aggregation functions like \textit{COUNT},  \textit{GROUP BY} and \textit{HAVING}. 
The example from Listing~\ref{lbl:ex1} defines the condition described above for all vehicles. To shorten the example, we omit the definition of prefixes for the queries.

{
	\small
	\begin{lstlisting}[language=JAVA, caption={Requirements Definition with 2 queries},captionpos=b,label=lbl:ex1,abovecaptionskip=15pt]
	//get all cars
	q1 = SELECT DISTINCT(COUNT(?car) as ?Cars)
	WHERE { ?car rdf:type ex:Car}		
		
	//cars with at least one engine
	q2 = SELECT ?car
	WHERE { 
		?car rdf:type ex:Car.
		?engine a ex:engine.
		?car ?connection ?engine.}
	GROUP BY ?car
	HAVING (COUNT(?engine) >= 1)
		
	//requirement: 'Each car has at least one engine'
	r1 : q1 = q2
\end{lstlisting}
}

As we can see, a comparison is made between the number of all cars in the solution model with those that actually have at least one engine.
If this number matches, the constraint is fulfilled.

There are a variety of different conditions that can be specified for a model. 
To express constraints, we have developed a specific grammar. An example is shown in Listing~\ref{lbl:grammar}.
It allows to link several conditions in a logical expression, so that complex constraints can be created based on the requests.
%
%
Besides using query results, static values can also be used. 
Furthermore, it is possible to make calculations based on the values and then evaluate them within a constrained.
This may be necessary for complex systems, where a large number of parameters influence the configuration of a solution model and possibly the optimal solution for a problem is searched for.
Finally, every constraint returns the value 'true' or 'false'.

{

%


\subsection{Solution Model Creation}
The solution model is created using the reference ontology.
For later verification with the RM and the constraints, it is important that there is a link between the reference and the solution.
For this purpose, an ontology is created which contains the constructed model in the form of instances.
First, it uses the instances that were specified by the RM itself.
For all newly created instances it is mandatory to assign the corresponding classes from the reference ontology to sustain the connection.

	\small
	\begin{lstlisting}[caption={Grammar Excerpt for Requirements Definition},captionpos=b,label=lbl:grammar,abovecaptionskip=10pt]
	Requirement ::=
	  'Not' (Requirement | SparqlQuery_Ask) |
	  Requirement LogicalExpression Requirement |
	  ComparisonFunction
	LogicalExpression ::=
	  'AND' | 'OR' | 'XOR' | '->'
	ComparisonFunction ::=
	  Value RelationalOperator Value
	RelationalOperator ::=
	  '<' | '<=' | '=' | '>=' | '>' | '!='

	Value ::=
	  FixedValue | Calculation | SparqlQuery
	...
\end{lstlisting}}

\begin{figure}[b]
  \centering
  \subfloat[][]{\includegraphics[width=0.41\linewidth]{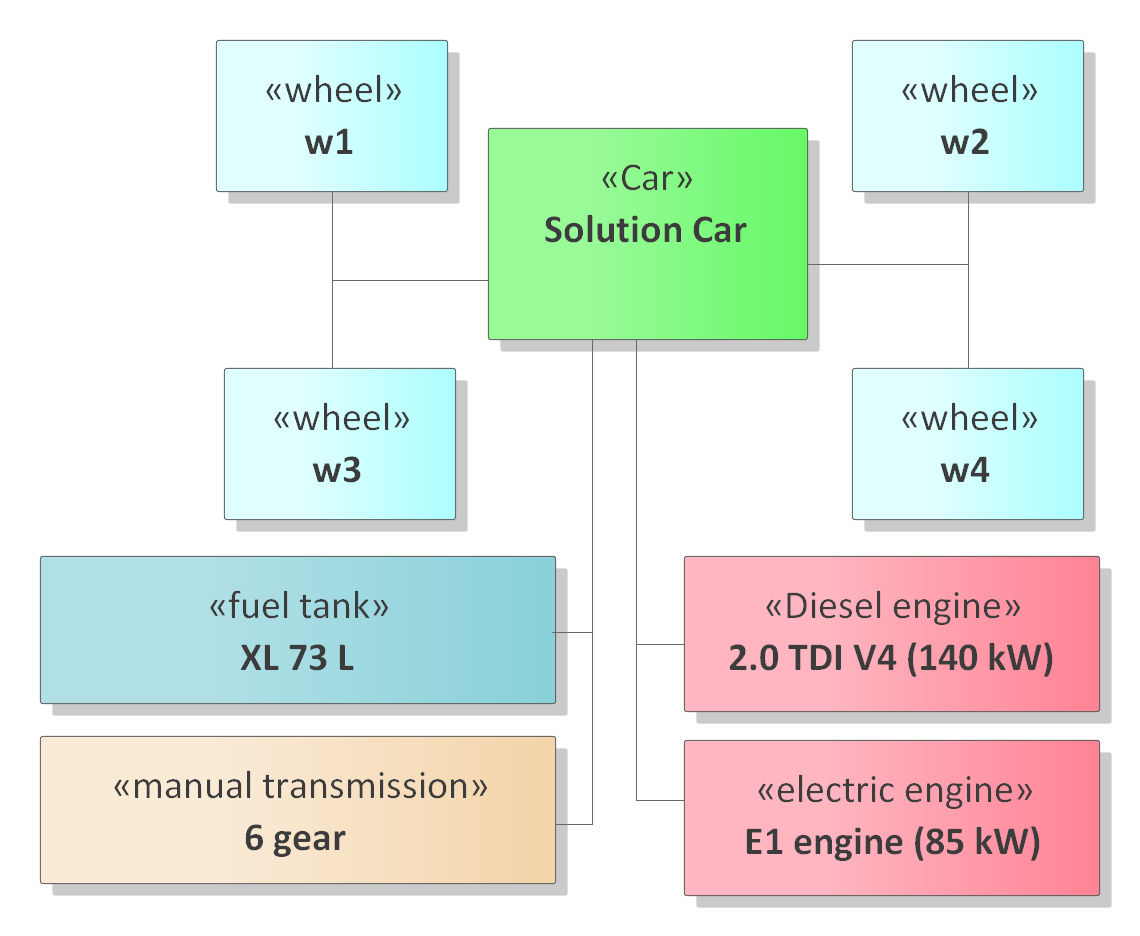}}%
  \qquad
  \subfloat[][]{\includegraphics[width=0.41\linewidth]{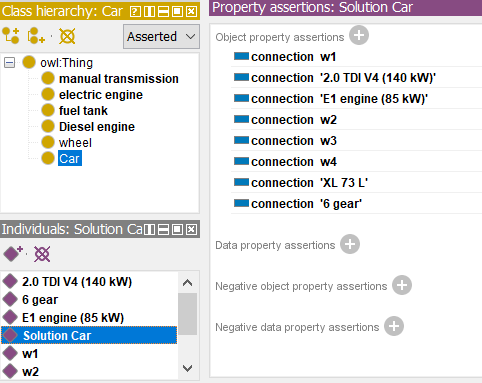}}%
  \caption{Solution model (a) and ontology (b) for the car example}%
  \label{fig:solution-model-ontology}
\end{figure}

Figure~\ref{fig:solution-model-ontology} (a) shows an example of a vehicle configuration as a UML model.
The assignment of a type is realized by stereotypes from the reference elements.
Other implementation strategies are also possible here.
This depends on the defined mapping between the modeling language and RDF.

The resulting ontology is shown in Figure~\ref{fig:solution-model-ontology} (b).
Thereby the classes obtain the same URIs as in the reference ontology.
This allows us to verify the correct merge of the building blocks.

\subsection{Model Verification and Validation}
The final step of our approach includes the verification of the solution model based on the constraints associated with the RM.
This verification is performed on the RDF representations of the models. 
An aligned ontology is generated by merging the reference ontology and the solution.
Using an RDFS reasoner, the knowledge resulting from the transitivity of the subclass relationship is inferred.
This ensures that instances of a subclass are also instances of the superordinate classes. 

Finally, the constraints are verified.
Therefore, the defined queries are executed on the merged ontology and the results are evaluated accordingly.
The result is a listing of all conditions with their corresponding truth values and the retrieved values.

For our RM, 11 queries were formulated. 
The queries and the requirements defined in Section~\ref{sec:problem} would be evaluated as shown in Table~\ref{tab:results}.
%
%
\begin{table}[]
\begin{tabular}{lr}
	\begin{tabular}{c| p{6cm}|c}
		\hline
		q1 &  \textit{COUNT:} all cars & 1 \\
		\hline
		q2 & \textit{COUNT:} cars with 4 wheels & 1 \\
		\hline
		q3 & \textit{COUNT:} cars with 1 transmission gear & 1 \\
		\hline
		q4 &  \textit{ASK:} transmission is a manual and an automatic transmission & false \\
		\hline
		q5 & \textit{ASK:} cars with less than one engine & false \\
		\hline
		q6 & \textit{ASK:} cars with more than one combustion engine & false  \\
		\hline
		q7 & \textit{COUNT:} cars with at least one combustion engine & 1 \\
		\hline
		q8 & \textit{COUNT:} cars with at least one electric engine  & 1 \\
		\hline
		q9 & \textit{COUNT:} cars with a combustion engine and a fuel tank & 1 \\
		\hline
		q10 & \textit{COUNT:} cars with an electric engine and a battery & 0 \\
		\hline
		q11 &\textit{COUNT:} cars with an electric engine and a automatic transmission & 0 \\
		\hline
	\end{tabular}
	&
	\begin{tabular}{|c|l|c|}
		\hline
		r1 & q1 = q2 & true \\
		\hline
		r2 & (q1 = q3) AND NOT q4 & true \\
		\hline
		r3 & NOT(q5 AND q6) & true \\
		\hline
		r4 &(q7=q9) AND (q8=q10) &false\\
		\hline
		r5 &  q8=q11 &  false  \\
		\hline
	\end{tabular}
	
\end{tabular}
\vspace*{0.3cm}
\caption{Queries and constraints with results for the car example}
\label{tab:results}
\end{table}

\section{Evaluation}
\label{sec:evaluation}

\begin{figure}[]
	\centering
	\includegraphics[width=.8\linewidth]{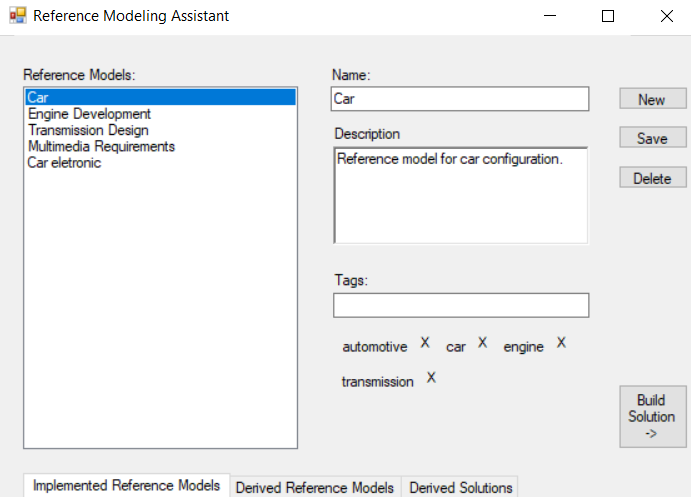}
	\caption{Reference Modeling Assistant}
	\label{fig:rma1}
\end{figure}

For the validation of our concept we designed a prototype as a standalone application in C\#, called Reference Modeling Assistant (RMA).
Figure~\ref{fig:rma1} shows an overview of the different RM.
Besides the name and the description, it is possible to assign tags to the models in order to facilitate the search for suitable models~\cite{Hillmann2022}.
Thereby, a RM can implement other RMs.
Thus, a multiple inheritance can be realized, whereby the constraints are taken over.
For example, a RM can exist for the general modeling of vehicles and other models that are specifically designed for the development of electric or combustion vehicles.
The general constraints then apply to both models.

With RMA it is possible to create simple ontologies using \textit{dotNETRDF}\footnote{https://www.dotnetrdf.org}, an open source .NET library for working with RDF triples.
An import of existing \textit{rdf} or \textit{owl} files is possible as well.

\begin{figure}[]
	\centering
	\includegraphics[width=0.8\linewidth]{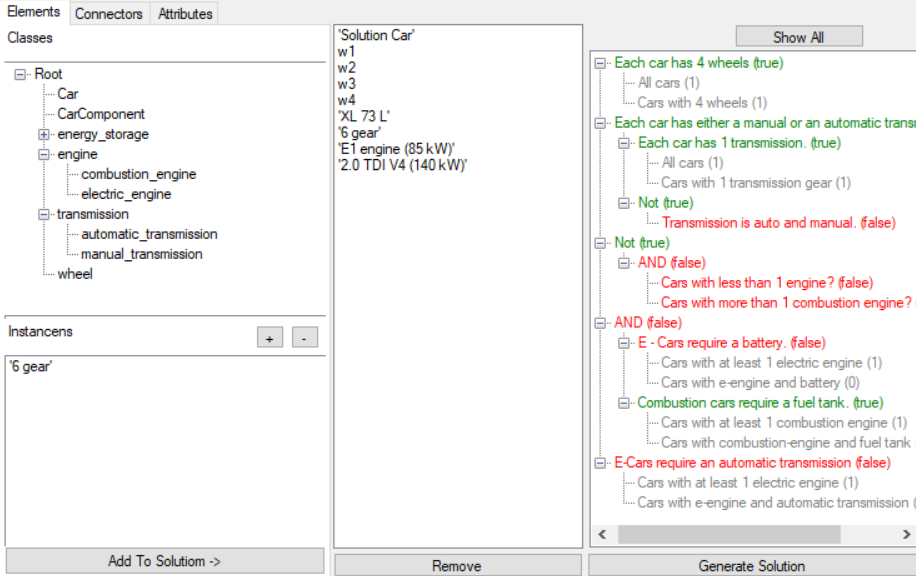}
	\caption{Constraints evaluation for the solution model}
	\label{fig:rma2}
\end{figure}

Once the RM have been constructed, a solution model can be created on their basis or an existing model can be validated against them.
Figure~\ref{fig:rma2} shows the \textit{Solution Builder}.
On the left side it displays the taxonomy with the classes of the RM and below the related instances.
The other two tabs give a choice for the available relations and attributes.

The middle column shows the created solution model, as we already know it from Figure~\ref{fig:solution-model-ontology}.
Depending on which tab is selected, the respective items can be added, removed and edited.

On the right side the constraints of the RM are shown.
These are validated directly when the solution is changed.
We see the queries listed in Table~\ref{tab:results} with their results and the constraints constructed from them.
The tree view shows the nesting of the logical statements.
A constrained is fulfilled when its upper node is displayed in green.
Unfulfilled constraints (4 and 5) are shown in red and queries which only returned values are shown in grey.

On the upper level, this representation allows a quick check of the extent to which a solution conforms to the constraints of a RM.
On the lower levels, the details of a query can be analyzed and appropriate corrections can be made to the model.

The result can be exported from the RMA as RDF file.
Afterwards, the content can be integrated into the individual modeling environment via transformation.

To demonstrate this step, we have developed an Add-In for the modeling tool \textit{Sparx Enterprise Architect (EA)}.
This allows the conversion of EA models as shown in Figure~\ref{fig:car} and~\ref{fig:solution-model-ontology} to RDF models and vice versa.

With the representation of the RM as an ontology a cross-language provision of a semantically enriched vocabulary is achieved.
The concepts and instances described within can be used to generate multiple solution models across different modeling languages.
It has been observed that the form of the constraints is strongly coupled to the modeling language and frameworks used.

The formalization of a RM requires some effort.
This is usually only profitable if the RM is reused often enough like in our example, where RMs are used to configure multiple vehicles.
Hence, the use of this approach is especially suitable for recurring modeling tasks and for those where high demands are made on model verification.

\section{Conclusion and Future Work}
\label{sec:conclusion_future_work}
In this paper, we presented a reference model design methodology. It begins with requirements engineering using RDF. In the process, it offers different possibilities to formal describe specific conditions. The reference models can be described and formalized into an ontology.
This reference design is used for the development of solution models. For this purpose, different building blocks are offered. In particular, various methods can be used to derive a specific solution from the reference design. The process of reuse consists beside others of the adaptation, adoption and extension of the reference building blocks. We have taken best practices into account and embedded them in a process with formalization.
Finally, our approach enables the subsequent verification and validation of the developed designs against the reference models for conformity.
Our reference modeling assistant highlight the interdependency. Thus, it finally contributes to compatibility between designs and quality assurance in context of maturity model.

In the future, we will establish a more integrated tool chain for the usability of reference models. Beside this, we are working on a structured repository of reference models and their easy retrieval as well as adaptability. Nevertheless, we will establish our approach in the context of system of system designs.

\bibliography{sample-ceur}

\begin{thebibliography}{27}
\expandafter\ifx\csname natexlab\endcsname\relax\def\natexlab#1{#1}\fi
\providecommand{\url}[1]{\texttt{#1}}
\providecommand{\href}[2]{#2}
\providecommand{\path}[1]{#1}
\providecommand{\DOIprefix}{doi:}
\providecommand{\ArXivprefix}{arXiv:}
\providecommand{\URLprefix}{URL: }
\providecommand{\Pubmedprefix}{pmid:}
\providecommand{\doi}[1]{\href{http://dx.doi.org/#1}{\path{#1}}}
\providecommand{\Pubmed}[1]{\href{pmid:#1}{\path{#1}}}
\providecommand{\bibinfo}[2]{#2}
\ifx\xfnm\relax \def\xfnm[#1]{\unskip,\space#1}\fi
\bibitem[{Hillmann et~al.(2022)Hillmann, Schnell, Hagel, and
  Karcher}]{Hillmann2022}
\bibinfo{author}{P.~Hillmann}, \bibinfo{author}{D.~Schnell},
  \bibinfo{author}{H.~Hagel}, \bibinfo{author}{A.~Karcher},
\newblock \bibinfo{title}{{Enterprise Model Library for
  Business-IT-Alignment}},
\newblock \bibinfo{journal}{Conference on Computer Science, Engineering and
  Applications}  (\bibinfo{year}{2022}).
\bibitem[{{The Open Group}(2022{\natexlab{a}})}]{IT4IT}
\bibinfo{author}{{The Open Group}}, \bibinfo{title}{{IT4IT}{\texttrademark}
  Standard, Version 3.0}, \bibinfo{type}{Technical Report}, The Open Group,
  \bibinfo{year}{2022}{\natexlab{a}}. \URLprefix
  \url{https://pubs.opengroup.org/it4it/3.0/standard/}.
\bibitem[{{The Open Group}(2022{\natexlab{b}})}]{Archimate32}
\bibinfo{author}{{The Open Group}}, \bibinfo{title}{ArchiMate{\textregistered}
  3.2 Specification}, \bibinfo{type}{Technical Report}, The Open Group,
  \bibinfo{year}{2022}{\natexlab{b}}. \URLprefix
  \url{https://pubs.opengroup.org/architecture/archimate3-doc/}.
\bibitem[{DIN SPEC 91345:2016-04(2016)}]{RAMI}
DIN SPEC 91345:2016-04, \bibinfo{title}{{Reference Architecture Model Industrie
  4.0} ({RAMI4.0})}, \bibinfo{year}{2016}.
\bibitem[{P\"ohn and Hillmann(2021)}]{Poehn2021}
\bibinfo{author}{D.~P\"ohn}, \bibinfo{author}{P.~Hillmann},
\newblock \bibinfo{title}{{Reference Service Model for Federated Identity
  Management}},
\newblock \bibinfo{journal}{International Conference on Evaluation and Modeling
  Methods for Systems Analysis and Development}  (\bibinfo{year}{2021}).
\bibitem[{Laurenzi et~al.(2018)Laurenzi, Hinkelmann, and van~der
  Merwe}]{Laurenzi.2018}
\bibinfo{author}{E.~Laurenzi}, \bibinfo{author}{K.~Hinkelmann},
  \bibinfo{author}{A.~van~der Merwe},
\newblock \bibinfo{title}{An agile and ontology-aided modeling environment},
\newblock in: \bibinfo{editor}{R.~A. Buchmann},
  \bibinfo{editor}{D.~Karagiannis}, \bibinfo{editor}{M.~Kirikova} (Eds.),
  \bibinfo{booktitle}{The Practice of Enterprise Modeling},
  \bibinfo{publisher}{Springer International Publishing},
  \bibinfo{address}{Cham}, \bibinfo{year}{2018}, pp. \bibinfo{pages}{221--237}.
\bibitem[{Glaser et~al.(2023)Glaser, Ali, Sallinger, and Bork}]{Glaser.2022}
\bibinfo{author}{P.-L. Glaser}, \bibinfo{author}{S.~J. Ali},
  \bibinfo{author}{E.~Sallinger}, \bibinfo{author}{D.~Bork},
\newblock \bibinfo{title}{Exploring enterprise architecture knowledge graphs
  in archi: The eakg toolkit},
\newblock in: \bibinfo{editor}{T.~P. Sales}, \bibinfo{editor}{H.~A. Proper},
  \bibinfo{editor}{G.~Guizzardi}, \bibinfo{editor}{M.~Montali},
  \bibinfo{editor}{F.~M. Maggi}, \bibinfo{editor}{C.~M. Fonseca} (Eds.),
  \bibinfo{booktitle}{Enterprise Design, Operations, and Computing. EDOC 2022
  Workshops}, \bibinfo{publisher}{Springer International Publishing},
  \bibinfo{address}{Cham}, \bibinfo{year}{2023}, pp. \bibinfo{pages}{332--338}.
\bibitem[{Karagiannis and Buchmann(2018)}]{Karagiannis2018APF}
\bibinfo{author}{D.~Karagiannis}, \bibinfo{author}{R.~A. Buchmann},
\newblock \bibinfo{title}{A proposal for deploying hybrid knowledge bases: the
  adoxx-to-graphdb interoperability case},
\newblock in: \bibinfo{booktitle}{Hawaii International Conference on System
  Sciences}, \bibinfo{year}{2018}. \URLprefix
  \url{https://api.semanticscholar.org/CorpusID:46946356}.
\bibitem[{Cinpoeru et~al.(2019)Cinpoeru, Ghiran, Harkai, Buchmann, and
  Karagiannis}]{Mihai.2019}
\bibinfo{author}{M.~Cinpoeru}, \bibinfo{author}{A.-M. Ghiran},
  \bibinfo{author}{A.~Harkai}, \bibinfo{author}{R.~A. Buchmann},
  \bibinfo{author}{D.~Karagiannis},
\newblock \bibinfo{title}{Model-driven context configuration in business
  process management systems: An approach based on knowledge graphs},
\newblock in: \bibinfo{editor}{M.~Pa{\'{n}}kowska},
  \bibinfo{editor}{K.~Sandkuhl} (Eds.), \bibinfo{booktitle}{Perspectives in
  Business Informatics Research}, \bibinfo{publisher}{Springer International
  Publishing}, \bibinfo{address}{Cham}, \bibinfo{year}{2019}, pp.
  \bibinfo{pages}{189--203}.
\bibitem[{Thomas and Fellmann(2009)}]{Oliver.2009}
\bibinfo{author}{O.~Thomas}, \bibinfo{author}{M.~Fellmann},
\newblock \bibinfo{title}{Semantic process modeling – design and
  implementation of an ontology-based representation of business processes},
\newblock \bibinfo{journal}{Business \& Information Systems Engineering}
  \bibinfo{volume}{1} (\bibinfo{year}{2009}) \bibinfo{pages}{438--451}.
  \DOIprefix\doi{10.1007/s12599-009-0078-8}.
\bibitem[{Becker and Delfmann(2004)}]{Becker.2004}
\bibinfo{editor}{J.~Becker}, \bibinfo{editor}{P.~Delfmann} (Eds.),
  \bibinfo{title}{{Referenzmodellierung: Grundlagen, Techniken und
  dom{\"a}nenbezogene Anwendung}}, \bibinfo{publisher}{Springer},
  \bibinfo{address}{Heidelberg}, \bibinfo{year}{2004}.
\bibitem[{Becker et~al.(2007)Becker, Delfmann, and Knackstedt}]{Becker.2007b}
\bibinfo{author}{J.~Becker}, \bibinfo{author}{P.~Delfmann},
  \bibinfo{author}{R.~Knackstedt},
\newblock \bibinfo{title}{{Adaptive Reference Modeling: Integrating
  Configurative and Generic Adaptation Techniques for Information Models}},
\newblock in: \bibinfo{editor}{J.~Becker}, \bibinfo{editor}{P.~Delfmann}
  (Eds.), \bibinfo{booktitle}{{Reference Modeling}},
  \bibinfo{publisher}{{Physica-Verlag HD}}, \bibinfo{address}{Heidelberg},
  \bibinfo{year}{2007}, pp. \bibinfo{pages}{27--58}.
\bibitem[{{Object Management Group (OMG)}(2014)}]{OCL24}
\bibinfo{author}{{Object Management Group (OMG)}}, \bibinfo{title}{Object
  Constraint Language. Version 2.4}, \bibinfo{type}{Technical Report}, {Object
  Management Group}, \bibinfo{year}{2014}. \URLprefix
  \url{https://www.omg.org/spec/OCL/2.4/}.
\bibitem[{Fettke and Loos(2006)}]{Fettke2006}
\bibinfo{author}{P.~Fettke}, \bibinfo{author}{P.~Loos},
  \bibinfo{title}{{Reference Modeling for Business Systems Analysis}},
  \bibinfo{publisher}{Idea Group Publishing}, \bibinfo{year}{2006}.
\bibitem[{Frank and Strecker(2007)}]{mci/Frank2007}
\bibinfo{author}{U.~Frank}, \bibinfo{author}{S.~Strecker},
\newblock \bibinfo{title}{Open reference models - community-driven
  collaboration to promote development and dissemination of reference models},
\newblock \bibinfo{journal}{Enterprise Modelling and Information Systems
  Architectures - An International Journal} \bibinfo{volume}{2}
  (\bibinfo{year}{2007}) \bibinfo{pages}{32--41}.
\bibitem[{{Signavio GmbH}(2020)}]{SignavioGmbH.2020}
\bibinfo{author}{{Signavio GmbH}}, \bibinfo{title}{{Business Transformation
  Suite}}, \bibinfo{year}{2020}. \URLprefix
  \url{https://www.signavio.com/de/products/business- transformation-suite/}.
\bibitem[{{Camunda Inc.}(2020)}]{CamundaInc..}
\bibinfo{author}{{Camunda Inc.}}, \bibinfo{title}{{Camunda BPM}},
  \bibinfo{year}{2020}. \URLprefix
  \url{https://camunda.com/de/products/camunda-bpm/}.
\bibitem[{Allemang et~al.(2005)Allemang, Polikoff, and Hodgson}]{Allemang2005}
\bibinfo{author}{D.~Allemang}, \bibinfo{author}{I.~Polikoff},
  \bibinfo{author}{R.~Hodgson},
\newblock \bibinfo{title}{{Enterprise Architecture Reference Modeling in
  OWL\/RDF}},
\newblock \bibinfo{journal}{International Semantic Web Conference}
  (\bibinfo{year}{2005}).
\bibitem[{Hooshmand et~al.(2017)Hooshmand, Adamenko, Kunnen, and
  K\"ohler}]{Hooshmand2017}
\bibinfo{author}{Y.~Hooshmand}, \bibinfo{author}{D.~Adamenko},
  \bibinfo{author}{S.~Kunnen}, \bibinfo{author}{P.~K\"ohler},
\newblock \bibinfo{title}{{An Approach for Holistic Model-Based Engineering of
  Industrial Plants}},
\newblock \bibinfo{journal}{Conference on Engineering Design (ICED)}
  (\bibinfo{year}{2017}).
\bibitem[{Ascher et~al.(2022)Ascher, Heiland, Schnell, Hillmann, and
  Karcher}]{Ascher2022}
\bibinfo{author}{D.~Ascher}, \bibinfo{author}{E.~Heiland},
  \bibinfo{author}{D.~Schnell}, \bibinfo{author}{P.~Hillmann},
  \bibinfo{author}{A.~Karcher},
\newblock \bibinfo{title}{{Methodology for Holistic Reference Modeling in
  Systems Engineering}},
\newblock \bibinfo{journal}{International Conference on Perspectives in
  Business Informatics Research}  (\bibinfo{year}{2022}).
\bibitem[{Becker et~al.(2008)Becker, Fleischer, Knackstedt, and
  Stein}]{Becker2008}
\bibinfo{author}{J.~Becker}, \bibinfo{author}{S.~Fleischer},
  \bibinfo{author}{R.~Knackstedt}, \bibinfo{author}{A.~Stein},
\newblock \bibinfo{title}{{Ontology Support for Configurative Reference
  Modeling}},
\newblock \bibinfo{journal}{European Conference on Information Systems (ECIS )}
   (\bibinfo{year}{2008}).
\bibitem[{Esswein(2016)}]{Esswein.2016}
\bibinfo{author}{W.~Esswein},
\newblock \bibinfo{title}{{Referenzmodelle und -modellierung}},
\newblock in: \bibinfo{editor}{T.~Benker}, \bibinfo{editor}{C.~J{\"u}rck},
  \bibinfo{editor}{M.~Wolf} (Eds.),
  \bibinfo{booktitle}{{Gesch{\"a}ftsprozessorientierte Systementwicklung}},
  volume~\bibinfo{volume}{46}, \bibinfo{publisher}{{Springer Vieweg}},
  \bibinfo{address}{Wiesbaden}, \bibinfo{year}{2016}, pp.
  \bibinfo{pages}{51--62}.
\bibitem[{Cyganiak et~al.(2014)Cyganiak, Wood, and Lanthaler}]{Cyganiak:14:RCA}
\bibinfo{author}{R.~Cyganiak}, \bibinfo{author}{D.~Wood},
  \bibinfo{author}{M.~Lanthaler}, \bibinfo{title}{{RDF} 1.1 Concepts and
  Abstract Syntax}, \bibinfo{type}{{W3C} Recommendation}, W3C,
  \bibinfo{year}{2014}. \URLprefix
  \url{https://www.w3.org/TR/2014/REC-rdf11-concepts-20140225/}.
\bibitem[{Seaborne and Harris(2013)}]{Seaborne:13:SQL}
\bibinfo{author}{A.~Seaborne}, \bibinfo{author}{S.~Harris},
  \bibinfo{title}{{SPARQL} 1.1 Query Language}, \bibinfo{type}{{W3C}
  Recommendation}, W3C, \bibinfo{year}{2013}. \URLprefix
  \url{https://www.w3.org/TR/2013/REC-sparql11-query-20130321/}.
\bibitem[{Kontokostas and Knublauch(2017)}]{Kontokostas:17:SCL}
\bibinfo{author}{D.~Kontokostas}, \bibinfo{author}{H.~Knublauch},
  \bibinfo{title}{Shapes Constraint Language ({SHACL})}, \bibinfo{type}{{W3C}
  Recommendation}, W3C, \bibinfo{year}{2017}. \URLprefix
  \url{https://www.w3.org/TR/2017/REC-shacl-20170720/}.
\bibitem[{Horrocks et~al.(2004)Horrocks, Patel-Schneider, Boley, Tabet,
  Grosofand, and Dean}]{Horrocks2004}
\bibinfo{author}{I.~Horrocks}, \bibinfo{author}{P.~F. Patel-Schneider},
  \bibinfo{author}{H.~Boley}, \bibinfo{author}{S.~Tabet},
  \bibinfo{author}{B.~Grosofand}, \bibinfo{author}{M.~Dean},
  \bibinfo{title}{{SWRL}: A Semantic Web Rule Language Combining {OWL} and
  {RuleML}}, \bibinfo{type}{Technical Report}, W3C, \bibinfo{year}{2004}.
  \URLprefix \url{http://www.w3.org/Submission/SWRL/}.
\bibitem[{Haudebourg and Tomaszuk(2023)}]{Haudebourg:23:RS}
\bibinfo{author}{T.~Haudebourg}, \bibinfo{author}{D.~Tomaszuk},
  \bibinfo{title}{{RDF} 1.2 Schema}, \bibinfo{type}{{W3C} Working Draft}, W3C,
  \bibinfo{year}{2023}. \URLprefix
  \url{https://www.w3.org/TR/2023/WD-rdf12-schema-20230615/}.

\end{thebibliography}

\end{document}